\documentclass[11pt,twoside]{article}
\usepackage{newpasp}

\usepackage{epsf}
\usepackage{graphicx}

\index{Mundell, C. G.}

\begin{document}

\title{The Gaseous Environment of Seyfert Galaxies}
\author{C. G. Mundell}
\affil{A.R.I., Liverpool John Moores University, Twelve Quays House, Egerton Wharf, Birkenhead, CH41 1LD, U.K.}

% A concise abstract is recommended.  Enter the text of the abstract in
% between the \begin{abstract} and \end{abstract} commands.  Do NOT
% include the word ``Abstract'' in your text; it is insterted
% automatically. Do NOT  make a paragraph break between \begin{abstract} 
% and the first line of the text of the abstract!  Abstracts are required 
% for all papers.

\begin{abstract}
The VLA has been used to image HI in a heterogeneous sample of nine
Seyferts as part of a continuing investigation of the role of galactic
host gas in the triggering and fuelling of nuclear activity.  Where
previous single-dish observations indicated complex, poorly understood
kinematics, improved angular resolution and sensitivity reveal tidal
tails and intragroup gas, distinguish dwarfs from main disks and
resolve non-linear gas dynamical features such as shocks in bars.
Seyferts may be abnormally gas-rich, but a larger, statistically
significant comparison of HI properties of normal and active galaxies
is now required.
\end{abstract}

% Include keywords if you wish. The keywords.apj file, found on aas.org 
% in the pubs/aastex-misc directory, contains a list of keywords used 
% with the ApJ and Letters.  

%\keywords{galaxies: Seyfert -- galaxies: gas -- galaxies: evolution}

% That's it for the front matter.  On to the main body of the paper.

% Here is a figure plotted using the \plotone command. 
% When plotting figures, keep in mind that the maximum width for
% the conferences series is 5.25in. 

% We could have also have used the \plotfiddle command
% \plotfiddle{file}{vsize}{rot}{horiz scale}{vert scale}{dx}{dy}
 
%% TO INCLUDE EQUATIONS OR TABLES, PLEASE SEE THE FILE
%% newpaspman.ps at http://www.aspsky.org/pubs/authors.html

\section{Introduction}

The triggering and continued fuelling of Active Galactic Nuclei (AGN)
is a key issue in astrophysics, but models involving release of
gravitational energy via accretion are incomplete in their description
of the interaction between the AGN and its environment.  While the
most luminous AGN, quasars, seem to coincide with violent dynamics in
the gas-rich universe at the epoch of galaxy formation (Haehnelt \&
Rees, 1993), nuclear activity in nearby galaxies is more problematic,
with reactivation of ubiquitous central black holes likely to
dominate.  Indeed, the rate of accretion or black-hole formation and
growth may be regulated by host galaxy properties, such as galactic
disk instabilities at different epochs (e.g., Shlosman \& Noguchi, 1993;
Sellwood, 1999).  The relative proximity of Seyfert galaxies makes
them ideal for the detailed study of both the AGN and its relationship
to the host galaxy environment.

\section{HI Studies of Seyfert Hosts}

Statistical studies of Seyfert hosts, conducted mainly at optical and
IR wavelengths show no conclusive links between nuclear activity and
host properties, such as bars or interactions - possible mechanisms
for triggering and fuelling activity. HI may be a better probe on
sub-pc to 100 kpc scales than the stellar component, often exhibiting
dramatic tidal disruption (e.g., NGC~3227 - Mundell et al., 1995) and
reacting non-linearly to even weak barred potentials (e.g. Mundell \&
Shone, 1999). Few detailed studies of HI in Seyferts have been
performed (Brinks \& Mundell, 1996), so we have embarked on a study to
investigate gas content, transport, distribution and activity-inducing
features such as bars and interactions.  Details of nine systems are
summarised in Table 1, with optical, HI and radio continuum images
shown in Figure 1.

\vspace*{8mm}
\begin{tabular}{cc}
\hspace*{-1.4cm}\textbf{\small Table 1. Summary of Seyfert host properties}\\
\hspace*{-1.2cm}
\begin{minipage}{5in}
\includegraphics[bb=100 280 490 605,angle=0,width=5.2in,clip=true]{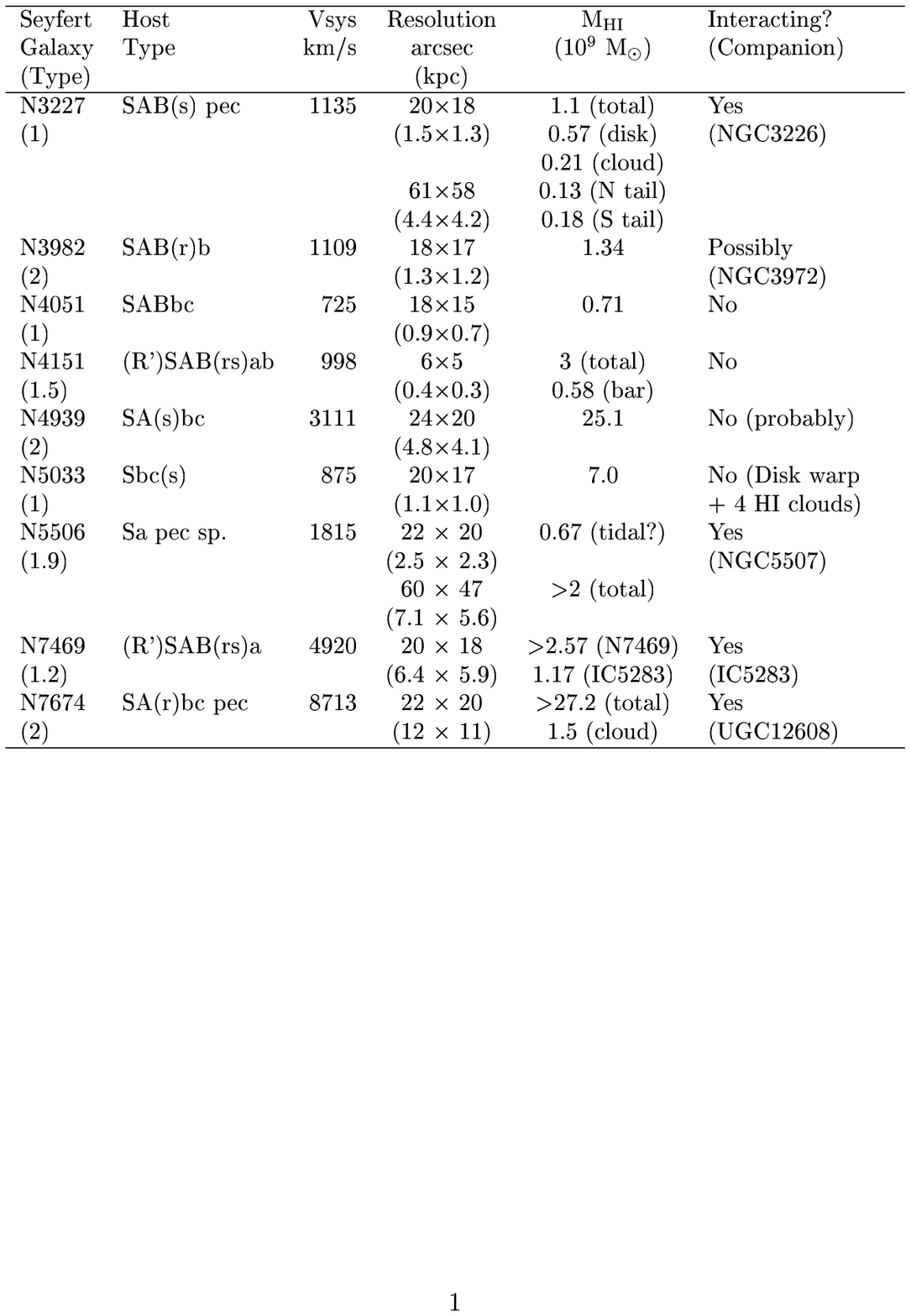}
\end{minipage}
\end{tabular}\\[1mm]

\subsection{The Need for Statistical Surveys}

Host-galaxy gas represents a reservoir of potential fuel for an AGN,
and given the ubiquity of supermassive black holes (Richstone et al.,
1998), the differences between normal and active galaxies must be more
closely related to the nature of the feeding rather than the presence
of a black hole.  A key question, therefore, is whether the gaseous
properties of normal galaxies differ from those with Seyfert
nuclei. Early results suggest that Seyfert hosts may be abnormally
gas-rich (Hunt et al., 1999) but a systematic, statistical comparison
is now required, achievable with a deep HI imaging survey of a matched
sample of Seyfert and normal galaxies.  This will have wider
application relevant to issues of galaxy formation and evolution,
e.g., the study of intragroup clouds and the issue of these as tidal
tails or relic material (e.g., see NGC~5506 in the Rogues Gallery this
volume). This gas would then provide tests of hierarchical structure
and galaxy formation, may explain Ly$\alpha$ clouds seen towards
quasars (e.g., Blitz et al., 1999) and possibly provide a new AGN
fuelling mechanism.

%\begin{figure}[t!]
%\plotfiddle{Huge.ps}{115mm}{0}{80}{80}{-210}{-130}
%\plotone{Huge.ps}
%\caption{\small Optical (DSS), HI and radio continuum image of nine
%Seyferts (see also Rogues Gallery - these proceedings).}
%\label{fig1}
%\end{figure}
%\vspace{-1cm}

\vspace{1.2cm}
\noindent
\hspace{2cm}
\begin{tabular}{cc}
\begin{minipage}{5in}
\includegraphics[bb=36 114 575 680,angle=0,width=4.25in,clip=true]{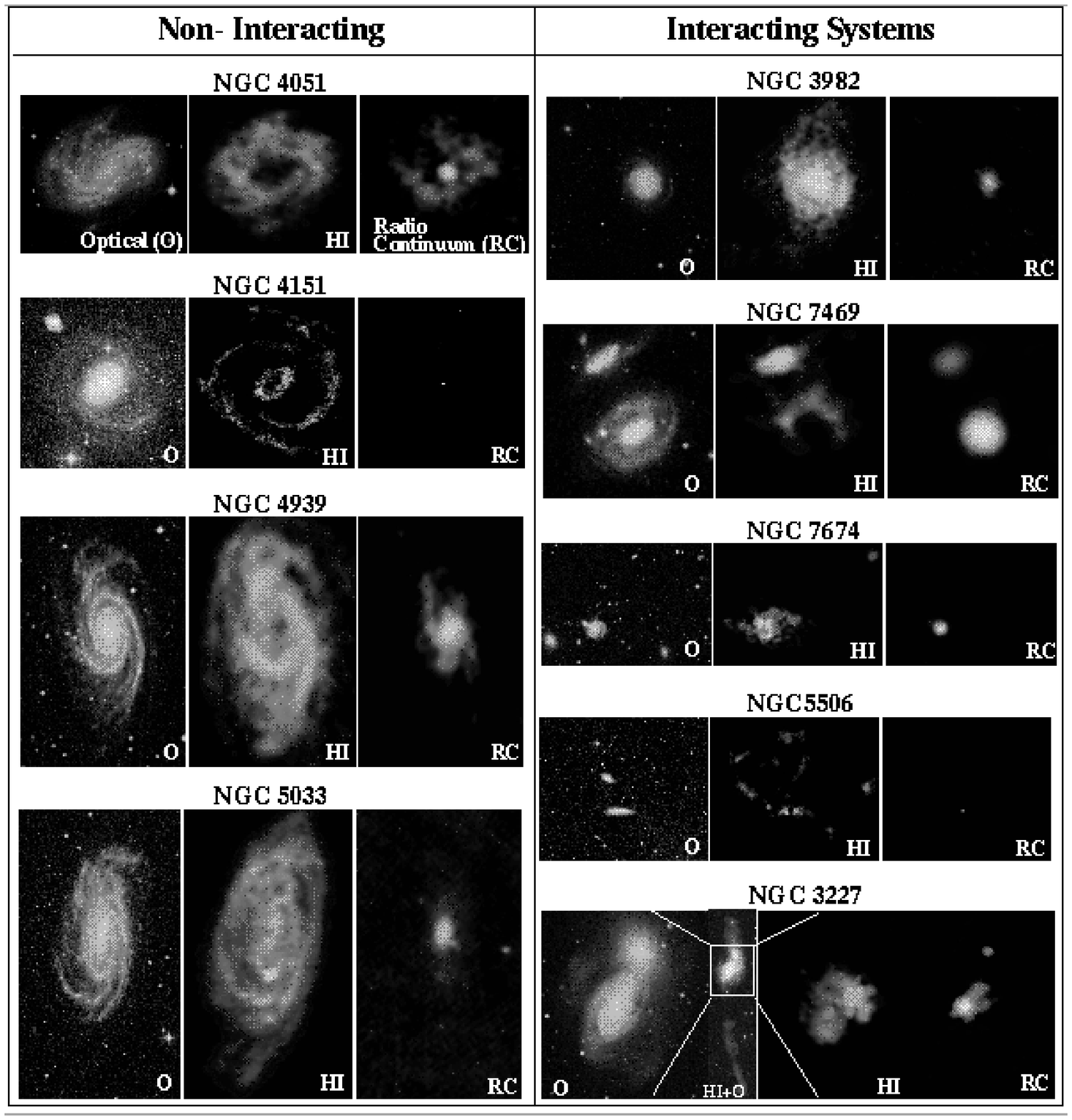}\\
\hspace*{0 cm}
\textbf{\small Figure 1. Optical (DSS2), HI and radio continuum images}
%\label{fig1}
\end{minipage}
\end{tabular}

\end{document}